\documentstyle[titlepage,12pt]{article}
\newcommand{\plb}[3]{Phys. Lett. {\bf B#1}, #3, (#2)}
\newcommand{\prl}[3]{Phys. Rev. Lett. {\bf #1}, #3, (#2)}
\newcommand{\prd}[3]{Phys. Rev. {\bf D#1}, #3, (#2)}
\newcommand{\npb}[3]{Nucl. Phys. {\bf B#1}, #3, (#2)}

\begin{document}
\begin{titlepage}
\rightline{}
\rightline{}
\rightline{}

\vspace*{2cm}

\addtocounter{footnote}{1}

\begin{center}

{\large \bf 
A comment on the index of the lattice Dirac operator \\
\vspace{0.5cm}
and the Ginsparg-Wilson relation. }

\vspace{3cm}

{\sc Atsushi Yamada}

\vspace{.5cm}

{\it Department of Physics, University of Tokyo, Tokyo, 113 Japan}

\vspace{2cm}

{\bf ABSTRACT}

\end{center}

We pursue Ginsparg and Wilsons' block spin approach in the derivation 
of the Ginsparg-Wilson relation and 
study the correspondence of the eigenmodes of the Dirac operators 
in the continuum and lattice theories. 
After introducing a suitable cut-off in the continuum theory, we identify    
unphysical modes of the lattice Dirac 
operator which do not correspond to any physical modes of the regulated 
continuum Dirac operator. We also consider zero modes in the continuum and 
lattice theories. Our studies give a physical 
interpretation of the expression of the index defined on a lattice 
and a formal argument on the relation of the indices between 
the continuum and lattice theories. 

\end{titlepage}


One of the recent developments in the treatment of the chiral symmetry 
on a lattice 
based on the Ginsparg-Wilson relation \cite{gw}  
\begin{eqnarray} 
 D^{lat} \gamma_{5} + \gamma_{5} D^{lat} = 
a D^{lat}\gamma_{5}D^{lat}
\label{eqn:gwori}
\end{eqnarray}
is the interesting index relation \cite{ha,lu}
\begin{eqnarray} 
tr^{lat} \gamma_{5}(1 -\frac{a}{2} D^{lat}) =n^{lat}_{+}-n^{lat}_{-} 
\label{eqn:inori}
\end{eqnarray}
on a lattice and its implications for the understanding of the 
anomaly \cite{gw}-\cite{adams}.    
In these equations, $D^{lat}$ is a Dirac operator describing a 
fermion on a lattice, $n^{lat}_{\pm}$ are the numbers of the right-handed and 
left-handed zero modes of 
$D^{lat}$, $a$ is the lattice spacing and $tr^{lat}$ 
is the trace in the lattice theory.  
The role of the factor $(1 -\frac{a}{2} D^{lat})$ in 
Eq. (\ref{eqn:inori}), which is absent in the continuum index 
relation \cite{ji,fujipath}
\begin{eqnarray} 
tr^{c} \gamma_{5}=n^{c}_{+}-n^{c}_{-},  
\label{eqn:inc}
\end{eqnarray}
is further studied in Refs. \cite{chiu,fujilat} based on the 
representation of the algebra (\ref{eqn:gwori}), 
leading to the observation 
that the mismatch of the chiralities between the zero modes of 
$D^{lat}$ should be compensated by the mismatch of the chiralities 
of its eigenmodes with the eigenvalue $\frac{2}{a}$ to ensure 
the relation $tr^{lat} \gamma_{5}=0$ in the lattice theory.   

In this short note, we pursue Ginsparg and Wilsons' block spin 
approach in the derivation of the Ginsparg-Wilson relation 
\cite{gw} and study the correspondence between the eigenmodes of a 
continuum Dirac operator $D^c$ and those of the  
lattice Dirac operator $D^{lat}$ constructed from $D^c$ 
following the block spin transformation, in the hope that such analyses will 
clarify understandings of the eigenmodes of $D^{lat}$ from a 
physical point of view.  
The eigenmodes of $D^{lat}$ with the eigenvalue $\frac{2}{a}$ and 
the zero modes of $D^c$ and $D^{lat}$ are investigated after 
introducing a suitable cut-off in $D^c$ to make our analysis  
free from divergences. This cut-off procedure, which was not 
considered in Ref. \cite{gw}, is an important step to derive 
a clear correspondence of the eigenmodes in our study.  
We will see that 
the eigenmodes of $D^{lat}$ with the eigenvalue $\frac{2}{a}$ do not 
correspond to any physical modes of $D^c$, thus they are  
considered to be unphysical \cite{c1}. 
Based on this criterion of the unphysical modes,  
we interpret that the role of the factor 
$(1- \frac{a}{2} D^{lat})$ in the 
$tr^{lat} \gamma_{5}(1 -\frac{a}{2} D^{lat}) $  
is to ensure that the unphysical modes $\lambda_n$  
satisfying $ D^{lat} \lambda= \frac{2}{a} \lambda $ are omitted 
in the evaluation of the trace.  
Also we will show that the zero modes of $D^c$ are transformed to 
the zero modes of $D^{lat}$ preserving the chirality so that 
$n^{c}_{\pm}=n^{lat}_{\pm}$. 
These two observations provide us a physical interpretation of the 
index expression (\ref{eqn:inori}) and the identity  
\begin{eqnarray} 
tr^{c}\gamma_{5}  
= tr^{lat} \gamma_{5}(1 -\frac{a}{2} D^{lat})
\label{eqn:index0}
\end{eqnarray}
at a formal level.  


We begin with an action $S^{c}(\bar{\phi}_x,\phi_x)$ of the fermionic fields 
$\phi_x$ and 
$\bar{\phi}_x$ defined 
in the continuum Euclidean space-time. 
From this action Ginsparg and Wilson constructed a new action 
$S^{lat}(\bar{\psi}_n,\psi_n)$ on a lattice by block spin transformation. 
First we define the block variables $\rho_n$ and 
$\bar{\rho}_n$ corresponding to the continuum fields $\phi_x$ and $\bar{\phi}_x$ as 
\begin{eqnarray} 
\rho_n=\sum_{x} f_{nx} \phi_x ,\,\,\,\, 
\bar{\rho}_n=\sum_{x} \bar{\phi}_x f^{*}_{xn}, 
\label{eqn:rho}
\end{eqnarray}
where the functions $f_{nx}$ and $f^{*}_{xn}$ have a sharp peak at $x=n$ and 
are proportional to the unit matrix in Dirac space.     
Using these variables, the new action $S^{lat}(\bar{\psi}_n,\psi_n)$ 
is defined by means of the block spin transformation \cite{block} as 
\begin{eqnarray} 
Ce^{- S^{lat}(\bar{\psi},\psi)} =  \int \prod_{x} d \bar{\phi}_x d\phi_x 
e^{- S^{c}(\bar{\phi},\phi)  -    
\alpha \sum_{n} (  \bar{\psi}_n - \bar{\rho}_n )  ( \psi_n - \rho_n ) },  
\label{eqn:block}
\end{eqnarray}
where $\alpha$ is a constant and will be equated to $2/a$ later.  
Assuming that $S^{c}(\bar{\phi},\phi)$ is quadratic in the fermion fields 
then so is $S^{lat}(\bar{\psi},\psi) $, 
and we may write 
\begin{eqnarray} 
S^{c}(\bar{\phi},\phi)= \sum_{xy} \bar{\phi}_x D^{c}_{xy} \phi_y,\,\,\,\,\,
S^{lat}(\bar{\psi},\psi)= \sum_{mn} \bar{\psi}_m D^{lat}_{mn} \psi_n. 
\label{eqn:dirac}
\end{eqnarray}
The Ginsparg-Wilson relation is a relation satisfied by 
$ D^{lat}_{mn} $ by virtue of its having 
been constructed from an initially chirally invariant action $S^{c}(\bar{\phi},\phi)$. 
The exponent of the right-hand side of Eq. (\ref{eqn:block}) is 
\begin{eqnarray} 
& & -\sum_{xy} \bar{\phi}_x D^{c}_{xy} \phi_y 
 -\alpha \sum_{n} (  \bar{\psi}_n - \bar{\rho}_n )  ( \psi_n - \rho_n )  
\nonumber \\
& & = 
\sum_{xy} \bar{\phi}_x \{  D^{c}_{xy} + \alpha \sum_{n}  f^{*}_{xn}  f_{ny}  \} \phi_y
+ \sum_{x} \bar{\phi}_x \xi_x + \sum_{y} \bar{\xi}_y \phi_y - 
\alpha \sum_{n} \bar{\psi}_n \psi_n,  
\label{eqn:z}
\\
& & \xi_x = \alpha \sum_{n}  f^{*}_{xn} \psi_n,\,\,\,\,\,  
\bar{\xi}_y = \alpha \sum_{n}  \bar{\psi}_n  f_{ny}. 
\nonumber        
\end{eqnarray}
Here we assume that our continuum theory is made well-defined by a 
certain regularization procedure and the spectrum of the 
Dirac operator 
$D^{c}_{xy}$ (namely the absolute value of its eigenvalues) 
is bounded from above. 
Therefore all the eigenmodes ${\lambda^{phys}_x }$ of 
the Dirac operator 
$D^{c}_{xy}$, which we want to simulate on a lattice, satisfy 
\begin{eqnarray}  
\sum_{y} D^{c}_{xy} \lambda^{i,phys}_y= \varepsilon_i \lambda^{i,phys}_x, 
\,\,\,\,\, | \varepsilon_i | \leq  \Lambda, 
\label{eqn:cut} 
\end{eqnarray}
for some cut-off $\Lambda$. 
Then for sufficiently large $\alpha$ (as will be seen later, 
$\alpha$ plays the role of the cut-off in the lattice theory), 
the operator $M_{xy}= D^{c}_{xy} + \alpha \sum_{n}  f^{*}_{xn}  f_{ny}  $ 
has no zero mode 
so that its inverse $M^{-1}_{xy}$ exists. 
For this choice of $\alpha$, the result of the 
path integral in Eq. (\ref{eqn:block}) is given by 
\begin{eqnarray} 
Ce^{- S^{lat}(\bar{\psi},\psi)} &=&  
det M e^{ 
\sum_{xy}  \bar{\xi}_x M^{-1}_{xy} {\xi}_y -\alpha \sum_{n}  
\bar{\psi}_n \psi_n   } 
\nonumber \\
&=&
det M e^{
- \sum_{mn}  \bar{\psi}_m 
\{  -\alpha^{2} \sum_{xy}  f_{mx} M^{-1}_{xy}  f^{*}_{yn} + 
\alpha \delta_{mn} \} 
{\psi}_n 
} 
\label{eqn:path1}
\end{eqnarray}
so we see \cite{com1}
\begin{eqnarray} 
D^{lat}_{mn} &=&   
\{  -\alpha^{2} \sum_{xy}  f_{mx} M^{-1}_{xy}  f^{*}_{yn} 
+ \alpha \delta_{mn} \}, 
\label{eqn:dlat}\\
& &M_{xy}=D^{c}_{xy}+ \alpha \sum_{n}  f^{*}_{xn}  f_{ny}.      
\label{eqn:m}
\end{eqnarray}


Now we consider the eigenmodes of $D^{lat}_{mn}$ 
and identify unphysical modes.  
Since the eigenvalues of $M_{xy}$ defined in Eq. (\ref{eqn:m}) for 
the eigenmodes $\lambda^{phys}_{x}$ are non-zero and finite, 
neither $\sum_{y} M_{xy} \lambda^{phys}_{y}$  
nor $\sum_{y} M^{-1}_{xy} \lambda^{phys}_{y}$ 
is equal to zero; 
\begin{eqnarray}
\sum_{y} M_{xy} \lambda^{phys}_{y} \neq 0,\,\,\,\,\,
\sum_{y} M^{-1}_{xy} \lambda^{phys}_{y} \neq 0.
\label{eqn:un}
\end{eqnarray} 
Therefore, a mode $\lambda_n $  of $D^{lat}_{mn}$ which satisfies 
\begin{eqnarray}
\sum_{n}  M^{-1}_{xy}  f^{*}_{yn}    \lambda_n = 0   
\label{eqn:unphysical}
\end{eqnarray}
is considered to be unphysical, because the eigenmode 
$\lambda_n$ on a lattice does not simulate a mode similar to 
any physical mode $\{ \lambda^{phys}_x  \}$. 
For $\lambda_n$ satisfying 
$\sum_{n}  M^{-1}_{xy}  f^{*}_{yn} \lambda_n=0$, 
we have 
\begin{eqnarray}
\sum_{n}  D^{lat}_{mn}    \lambda_n = \alpha   \lambda_n   
\label{eqn:unphysicaleg}
\end{eqnarray}
so the eigenmodes of $D^{lat}_{mn}$ with the eigenvalue $\alpha $ 
are considered to be unphysical on a lattice, because they have no 
counterparts among the physical spectrum $\{\lambda^{phys}_{y} \} $ of 
$D^c_{xy}$. 


Next we consider a zero mode $\lambda^{0,phys}_x$ of 
$D^{c}_{xy}$. The eigenvalue equation   
$\sum_{y} D^{c}_{xy} \lambda^{0,phys}_y=0$ 
and Eq. (\ref{eqn:m}) yield   
\begin{eqnarray}
\sum_{y} M_{xy}  \lambda^{0,phys}_y 
= \sum_{y} \alpha \sum_{n}  f^{*}_{xn}  f_{ny}\lambda^{0,phys}_y.  
\end{eqnarray}
Since $M^{-1}_{xy}$ exists, multiplying $\sum_{z} f_{mz} M^{-1}_{zx} $ 
for both sides of 
the above equation, we obtain  
\begin{eqnarray}
& &\lambda^{0,phys}_m = 
\alpha \sum_{z,x,n} f_{mz} M^{-1}_{zx} f^{*}_{xn} \lambda^{0,phys}_n 
\label{eqn:zero1}, \\
& &\lambda^{0,phys}_m = \sum_{x} f_{mx}\lambda^{0,phys}_x. 
\label{eqn:zero3}
\end{eqnarray}
Because of the identity (\ref{eqn:zero1}), we have  
\begin{eqnarray}
\sum_{n} D^{lat}_{mn}  \lambda^{0,phys}_n &=&  \sum_{n}
\{  -\alpha^{2} \sum_{xy}  f_{mx} M^{-1}_{xy}  f^{*}_{yn} 
+ \alpha \delta_{mn} \} \lambda^{0,phys}_n 
\nonumber \\
&=& -\alpha \lambda^{0,phys}_m + \alpha \lambda^{0,phys}_m =0,
\label{eqn:zero2}
\end{eqnarray}
therefore $\lambda^{0,phys}_n $  is a zero mode of $D^{lat}_{mn}$. 
The functions $\{f_{nx} \}$ are proportional to the unit matrix in Dirac space 
so that the chirality of $\lambda^{0,phys}_m $ is same as that of 
$\lambda^{0,phys}_x $. So all the zero modes of $D^{c}_{xy}$ have the 
corresponding zero modes of $D^{lat}_{mn}$ preserving their chiralities. 
This is in some sense desirable and naively expected result, 
and we have seen that it is in fact shown regardless of the 
details of the block spin transformation, e.g., 
the choice of the functions $f$ and $f^{*}$.    
Here we assume that our lattice is fine enough so that 
the shape of the zero modes $\lambda^{0,phys}_x$ is well preserved 
on a lattice when transformed to the zero modes $\lambda^{0,phys}_n$ 
of $D^{lat}_{mn}$ and they are distinguishable with each other 
on a lattice.


Our criterion of the unphysical modes on a lattice leads us to define 
the index on a lattice among the physical modes to be 
\begin{eqnarray} 
tr^{lat} \gamma_{5}(1 -\frac{1}{\alpha} D^{lat}). 
\label{eqn:index1}
\end{eqnarray} 
After eliminating the contribution of the unphysical modes by the factor 
$(1 -\frac{1}{\alpha} D^{lat})$, 
the index (\ref{eqn:index1}) is equal to 
$n^{lat}_+ - n^{lat}_-$ 
where $n^{lat}_+$ and $n^{lat}_-$ are the number of the 
right-handed and left-handed zero modes of $D^{lat}$ \cite{ha,lu}.       
Our analysis on the zero modes in the continuum and lattice 
theories naively yields $n^{c}_{\pm}= n^{lat}_{\pm}$, 
where $n^{c}_+$ and $n^{c}_-$ are the number of the 
right-handed and left-handed zero modes of $D^c$. 
These two observations give rise to the 
index relation among the physical modes as 
\begin{eqnarray} 
tr^{c}\gamma_{5}  
= tr^{lat} \gamma_{5}(1 -\frac{1}{\alpha} D^{lat}). 
\label{eqn:index3}
\end{eqnarray}

 
Here we derive the Ginsparg-Wilson relation starting 
from Eq. (\ref{eqn:block}) to see that $\alpha=2/a$ 
and the index relation (\ref{eqn:index3}), which was obtained by 
studying the correspondence of the eigenmodes, is also obtained 
as a by-product in the derivation. 
Under a global chiral transformation  
$\psi \rightarrow  e^{-i\epsilon \gamma_{5}}\psi$ and 
$\bar{\psi}  \rightarrow  \bar{\psi} e^{-i\epsilon \gamma_{5}} $, 
we have 
\begin{eqnarray} 
Ce^{- S^{lat}(\bar{\psi} e^{-i\epsilon \gamma_{5}} , 
e^{-i\epsilon \gamma_{5}}   \psi) } &=&  \int \prod_{x} d \bar{\phi}_x d\phi_x 
e^{- S^{c}(\bar{\phi},\phi)  -    
\alpha \sum_{n} (  \bar{\psi}_n e^{-i\epsilon \gamma_{5}}   - \bar{\rho}_n ) 
 ( e^{-i\epsilon \gamma_{5}}\psi_n - \rho_n ) }  
\nonumber \\
&=& 
e^{2i\epsilon tr^{c}\gamma_{5}}
\int \prod_{x}    d \bar{\phi}'_x d\phi'_x 
e^{- S^{c}(\bar{\phi}',\phi') -    
\alpha \sum_{n} (  \bar{\psi}_n    - \bar{\rho}'_n ) 
  e^{-2i\epsilon \gamma_{5}  }(\psi_n - \rho_n' ) }
\label{eqn:trans}
\end{eqnarray}
where the second line follows from a change of variables   
$\phi \rightarrow \phi'= e^{i\epsilon \gamma_{5}}\phi $ and 
$\bar{\phi}  \rightarrow  \bar{\phi}'= 
\bar{\phi} e^{i\epsilon \gamma_{5}} $, and ${\rho}'_n$ 
and $\bar{\rho}'_n$ are the block variables constructed from 
$ \phi'$ and  $\bar{\phi}'$. We have also kept the Jacobian factor 
$2i\epsilon tr^{c}\gamma_{5}$ 
of this change of variables, which exhibits some subtleties for the physical 
modes $\{ \lambda^{phys}_x \}$ of the Dirac operator 
$D^{c}_{xy}$ with gauge fields of topologically nontrivial 
nature \cite{fujipath}.  
The right-hand side of Eq. (\ref{eqn:trans}) is expanded 
to first order in $\epsilon$ as
\begin{eqnarray} 
& &(1+ {2i\epsilon tr^{c}\gamma_{5}} ) 
\int \prod_{x}    d \bar{\phi}'_x d\phi'_x 
e^{- S^{c}(\bar{\phi}',\phi')  -    
\alpha \sum_{n} (  \bar{\psi}_n    - \bar{\rho}'_n ) (\psi_n - \rho_n' ) }
\{ 1+ 2i\epsilon \alpha \sum_{n} 
(  \bar{\psi}_n    - \bar{\rho}'_n ) \gamma_{5}(\psi_n - \rho_n' ) 
\}
\nonumber \\
& &=
(1+  2i\epsilon tr^{c}\gamma_{5}  ) 
\int \prod_{x}    d \bar{\phi}'_x d\phi'_x 
\{ 1+ 2i\epsilon \alpha \sum_{n} 
(  \bar{\psi}_n    - \bar{\rho}'_n ) \gamma_{5}
(-\frac{1}{\alpha }) 
\frac{\partial}{\partial \bar{\psi}_n }
\}
e^{- S^{c}(\bar{\phi}',\phi')  -    
\alpha \sum_{n} (  \bar{\psi}_n    - \bar{\rho}'_n ) 
(\psi_n - \rho_n' ) }
\nonumber \\
& &=
(1+ {2i\epsilon tr^{c}\gamma_{5}} ) 
[ 1- 2i\epsilon \{ tr^{lat} \gamma_{5} + \frac{1}{\alpha} 
\frac{\partial}{\partial \psi_n }  \gamma_{5}
\frac{\partial}{\partial \bar{\psi}_n }
\}]
C e^{- S^{lat}(\bar{\psi},\psi)}
\nonumber \\
& &= 
[ 1+ 
  2i\epsilon \{ tr^{c}\gamma_{5}   
- tr^{lat} \gamma_{5}(1 -\frac{1}{\alpha} D^{lat}) \}  
+2i \epsilon \frac{1}{\alpha} \bar{\psi}D^{lat}\gamma_{5}D^{lat}\psi]
C e^{- S^{lat}(\bar{\psi},\psi)}
\end{eqnarray}
while the left-hand side is expanded as 
\begin{eqnarray} 
Ce^{- S^{lat}(\bar{\psi} e^{-i\epsilon \gamma_{5}} , 
e^{-i\epsilon \gamma_{5}}   \psi)}
\{ 1 + 
i \epsilon  \bar{\psi} ( D^{lat} \gamma_{5} + \gamma_{5} D^{lat}  )  \psi
\}.
\end{eqnarray}
Comparing these two expressions, we obtain the Ginsparg-Wilson relation 
\begin{eqnarray} 
 D^{lat} \gamma_{5} + \gamma_{5} D^{lat} = 
\frac{2}{\alpha} D^{lat}\gamma_{5}D^{lat}
\label{eqn:gw}
\end{eqnarray}
with $\alpha=2/a$ and the index relation 
\begin{eqnarray} 
2tr^{c}\gamma_{5}  
= 2tr^{lat} \gamma_{5}(1 -\frac{1}{\alpha} D^{lat}). 
\label{eqn:index}
\end{eqnarray}


Finally a few comments are in order. 
Our criterion of the unphysical modes of $D^{lat}$ is consistent 
with the definitions of the scalar and pseudo-scalar densities 
$S_m$ and $P_m$ 
proposed in Refs. \cite{ha2,n1,chan} 
\begin{eqnarray} 
S_m= \sum_{n}\bar{\psi}_{m}
(1 -\frac{1}{\alpha} D^{lat})_{mn} {\psi}_{n},
\,\,\,\,\,
P_m= \sum_{n}\bar{\psi}_{m}\gamma_{5}
(1 -\frac{1}{\alpha} D^{lat})_{mn} {\psi}_{n},
\label{eqn:sp}
\end{eqnarray}
where we interpret that the role of 
$ (1 -\frac{1}{\alpha} D^{lat})$ is to eliminate 
the contributions of the unphysical modes.

The precise relation between the high energy modes 
of $D^{c}$ and the high energy modes of 
$D^{lat}$ with the eigenvalue not equal to $\alpha$ 
depends on the choice of the cut-off scales $\Lambda$ 
in Eq. (\ref{eqn:cut}) and $\alpha$. 
However these high energy modes are always vector-like 
both in the continuum and lattice theories \cite{chiu,fujilat}  
and will decouple from the low energy physics. 
Thus detailed prescriptions of the ultraviolet cut-off 
do not change our observations.     
 

I would like to thank K.~Fujikawa for encouraging me to 
write this short note. 
I am also grateful to M.~Ishibashi for careful reading of the 
manuscript and useful comments, and to K.~Nagai for useful comments.


\begin{thebibliography}{99}
\bibitem{com1}
The Ginsparg-Wilson relation is also 
obtained from these two equations  
assuming that $D^{c}_{xy} \gamma_{5} + \gamma_{5}D^{c}_{xy} =0$.
\end{thebibliography}
\end{document}